\documentclass[letterpaper, 10 pt, conference]{ieeeconf}  

\IEEEoverridecommandlockouts                              
\usepackage{graphicx}
\usepackage{xcolor}
\usepackage[caption=false,font=footnotesize]{subfig}
\usepackage{url}
\usepackage{cite}

\usepackage{array, amsmath, amssymb, amsfonts, bm}
\usepackage{mathrsfs}
\usepackage{multirow, booktabs}
\usepackage{algorithm, algpseudocode}
\usepackage{xargs}

\usepackage{caption}
\usepackage{comment}
\usepackage{kantlipsum}

\long\def\symbolfootnote[#1]#2{\begingroup\def\thefootnote{\fnsymbol{footnote}}\footnote[#1]{#2}\endgroup}

\newcommand{\setR}{\mathbb{R}}
\newcommand{\setRp}{\mathbb{R}_+}
\newcommand{\setC}{\mathbb{C}}

\newcommand{\eye}{{\bf I}}

\newcommand{\T}{\mathsf{T}}
\newcommand{\adj}{\mathsf{H}}

\newcommand{\distlognormal}[2]{\mathcal{LN}\left({#1}, {#2}\right)}
\newcommand{\distcmpnormal}[2]{\mathcal{N}_{\mathbb{C}}\left({#1}, {#2}\right)}

\newcommand{\distcategorical}[1]{\mathrm{Cat}\left({#1}\right)}
\newcommand{\distdirichlet}[1]{\mathrm{Dir}\left({#1}\right)}

\newcommand{\E}{\mathbb{E}}

\newcommand{\KL}{\mathcal{D}_\mathrm{KL}}

\NewDocumentCommand\newletter{m m o m m}{
\NewDocumentCommand#1{s t@ o}{%
\IfBooleanTF{##1}{\mathbf{\MakeUppercase{#2}}\IfValueT{#3}{^{#3}}}{%
\IfBooleanTF{##2}{\mathbf{#2}\IfValueT{#3}{^{#3}}_{\IfValueTF{##3}{##3}{#5}}}{%
{#2}\IfValueT{#3}{^{#3}}_{\IfValueTF{##3}{##3}{#4}}%
}}}}

\NewDocumentCommand\newletterbm{m m o m m}{
\NewDocumentCommand#1{s t@ o}{%
\IfBooleanTF{##1}{\bm{\MakeUppercase{#2}}\IfValueT{#3}{^{#3}}}{%
\IfBooleanTF{##2}{\bm{#2}\IfValueT{#3}{^{#3}}_{\IfValueTF{##3}{##3}{#5}}}{%
{#2}\IfValueT{#3}{^{#3}}_{\IfValueTF{##3}{##3}{#4}}%
}}}}
\newletter{\y}{y}{ijc}{ij}
\newletter{\x}{x}{tfm}{tf}
\newletter{\s}{s}{tfk}{tf}
\newletter{\n}{n}{tfm}{tf}
\newletter{\sv}{a}{fkm}{fk}
\newletter{\svt}{b}{fdm}{fd}
\newletterbm{\psd}{\lambda}{tfk}{tf}

\newcommand{\scm}[1][fk]{\mathbf{H}_{#1}}
\newcommand{\scmt}[1][fd]{\mathbf{G}_{#1}}

\newcommand{\escm}[1][fk]{\hat{\mathbf{H}}_{#1}}

\newletter{\z}{z}{tfk}{tf}
\newletter{\w}{w}{kd}{k}
\newletter{\ew}{w}[*]{kd}{k}

\newletterbm{\zrate}{\pi}{k}{}
\newletterbm{\ezrate}{\pi}[*]{k}{}
\newletterbm{\wrate}{\phi}{kd}{k}
\newletterbm{\krate}{\psi}{k}{k}

\newcommand{\elbo}{\mathcal{L}}

\title{\LARGE \bf Self-supervised Neural Audio-Visual Sound Source Localization \\ via Probabilistic Spatial Modeling}

\author{Yoshiki Masuyama$^{1,2}$\!, Yoshiaki Bando$^1$\!, Kohei Yatabe$^2$\!, Yoko Sasaki$^1$\!, Masaki Onishi$^1$\!, Yasuhiro Oikawa$^2$%
\thanks{$^1$National Institute of Advanced Industrial Science and Technology, Japan
 {\tt\small \{masuyama.yoshiki, y.bando\}@aist.go.jp}}%
\thanks{$^2$Department of Intermedia Art and Science, Waseda University, Japan}
}

\begin{document}

\setlength\abovedisplayskip{1.3mm}
\setlength\belowdisplayskip{1.3mm}

\maketitle
\thispagestyle{empty}
\pagestyle{empty}

\begin{abstract}
Detecting sound source objects within visual observation is important for autonomous robots to comprehend surrounding environments.
Since sounding objects have a large variety with different appearances in our living environments, labeling all sounding objects is impossible in practice.
This calls for self-supervised learning which does not require manual labeling.
Most of conventional self-supervised learning uses monaural audio signals and images and cannot distinguish sound source objects having similar appearances due to poor spatial information in audio signals.
To solve this problem, this paper presents a self-supervised training method using $\bm{360^\circ}$ images and multichannel audio signals.
By incorporating with the spatial information in multichannel audio signals, our method trains deep neural networks (DNNs) to distinguish multiple sound source objects.
Our system for localizing sound source objects in the image is composed of audio and visual DNNs.
The visual DNN is trained to localize sound source candidates within an input image.
The audio DNN verifies whether each candidate actually produces sound or not.
These DNNs are jointly trained in a self-supervised manner based on a probabilistic spatial audio model.
Experimental results with simulated data showed that the DNNs trained by our method localized multiple speakers.
We also demonstrate that the visual DNN detected objects including talking visitors and specific exhibits from real data recorded in a science museum.

\end{abstract}

\section{Introduction}

Autonomous robots have to comprehend surrounding environments to decide their actions in our living circumstances.
Major tasks of those robots are human-robot interaction~\cite{action1} and security surveillance~\cite{security1}, which require localization and recognition of sound sources around the robots.
While detection of the direction of arrivals (DoAs) of the sound has been performed based mainly on audio information obtained by a set of microphones~\cite{music,phat,dsvd}, taking its correspondence to objects in the environment is also important for manipulating the robots.
This requires to find sounding objects within visual observation, which is referred to audio-visual sound source localization (AV-SSL) in this paper.

\begin{figure}
    \centering
    \vspace{-2mm}
    \hfill
    \subfloat[Robot in museum]{\includegraphics[height=2.8cm]{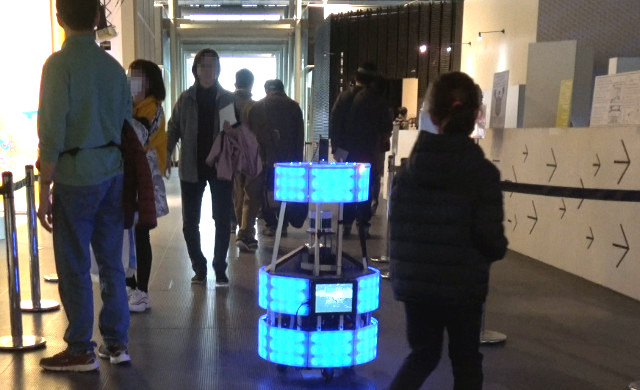}}
    \hspace{1mm}
    \subfloat[Head of robot]{\includegraphics[height=2.8cm]{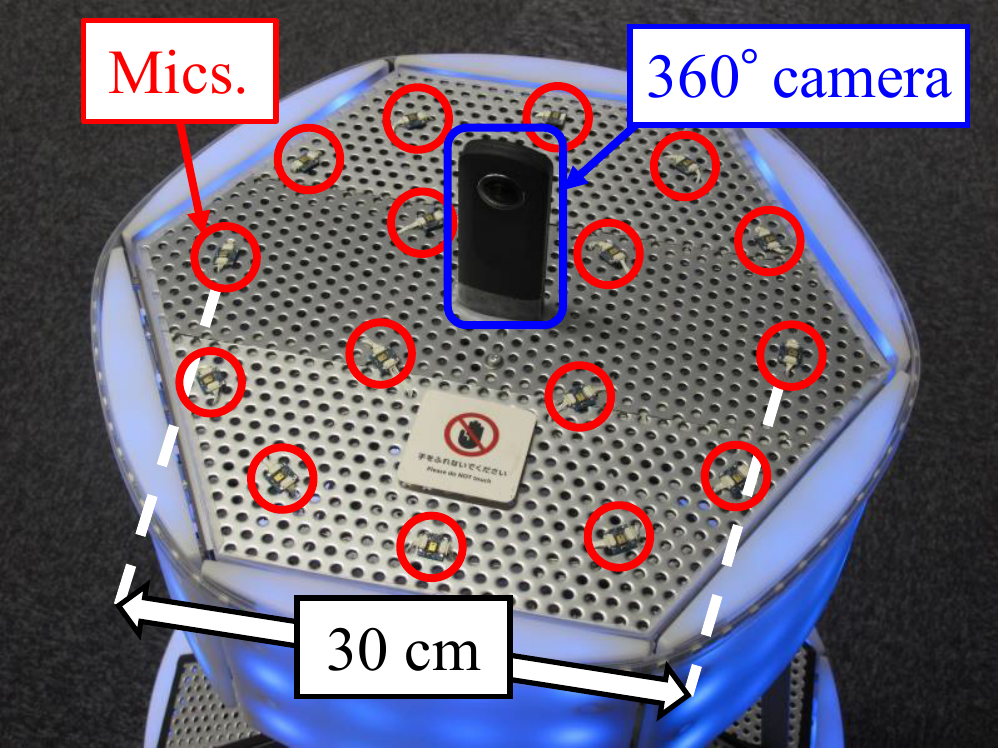}}
    \hfill
    \setlength\abovecaptionskip{0mm}
    \caption{Autonomous robot called Peacock.}
    \label{fig:peacock}
    \vspace{0mm}
    \includegraphics[width=0.9\hsize]{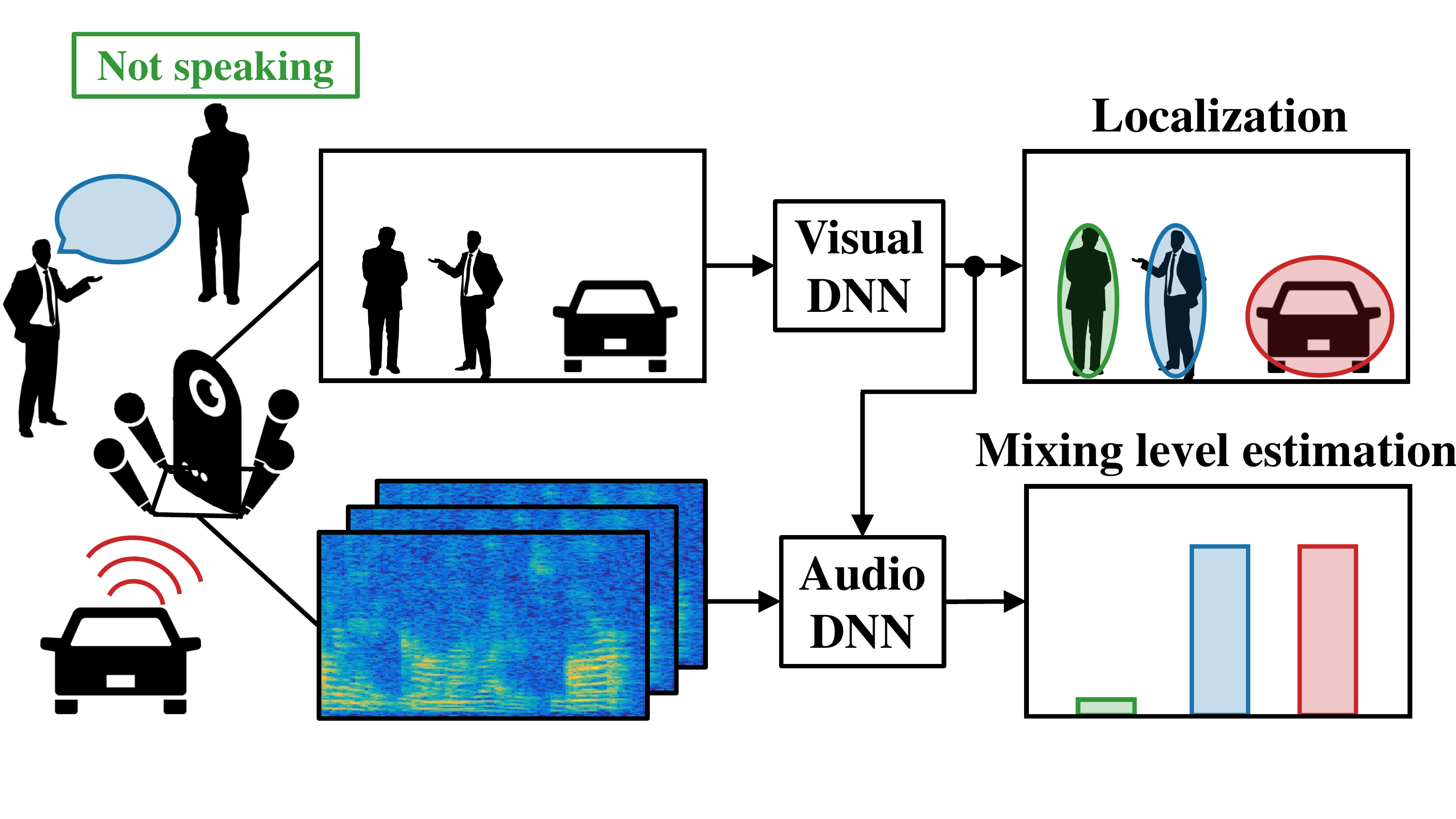}
    \setlength\abovecaptionskip{-3mm}
    \setlength\belowcaptionskip{-4mm}
    \caption{Illustration of our multichannel AV-SSL system: sound source localization with mixing level estimation.}
    \label{fig:problem}
\end{figure}

The difficulty of AV-SSL arises from the excess diversity of the sounding objects.
For instance, an autonomous robot interacting with an audience in a museum (as in Fig.~\ref{fig:peacock}) is exposed to sounds emitted from not only the target speaker but also the surrounding crowd and exhibits having very different appearances.
Since recording and labeling every sound source object (not restricted to human) with all possible sound and appearance are impossible in practice, a strategy without manual supervision is necessary for learning the complicated relation between audio and visual observations.

Self-supervised learning is one promising approach to AV-SSL as it does not require manual labeling.
In the existing literature~\cite{ssavssl1,ssavssl2,ssavssl3}, two deep neural networks (DNNs), an audio network and a visual network, are utilized to relate the same event observed in both of the audio and visual recordings.
They are trained so that their output features coincide with each other, where the data required for training is only monaural audio signals with the corresponding images~\cite{ssavssl1,ssavssl2} (or videos~\cite{ssavssl3}).
These methods can learn various kinds of sounding objects contained in the training data and perform well for audio events having different appearances (e.g., a musical performance with different instruments).
However, applying them to audio events consisting of multiple sound sources having similar appearances (e.g., people talking simultaneously, which is typical in our daily life) is not easy~\cite{ssavsslm}.
This should be because monaural audio signals do not contain spatial information which is a key to localize objects in visual observations.

In this paper, we present a self-supervised training method of audio and visual DNNs for AV-SSL using a $360^\circ$ camera and a microphone array as illustrated in Fig.~\ref{fig:problem}.
The proposed method is composed of two DNNs: visual DNN for object detection and audio DNN for source number estimation.
The visual DNN detects and localizes sound source candidates within the input image, where the candidates are allowed to include silent objects which should be ignored in AV-SSL.
Then, the audio DNN estimates the mixing levels (mixing proportion) of sound for the candidates to omit the silent objects as in Fig.~\ref{fig:problem}.
The key idea of our method is training both audio and visual DNNs based on a probabilistic spatial audio model, which allows us to train the two DNNs without manual labeling.
This training is formulated as a probabilistic inference of a spatial audio model that has DoAs and mixing levels of sound sources as latent variables.
In summary, the main contributions of this paper are as follows:
(1) proposing a self-supervised training method for AV-SSL using multichannel audio signals to distinguish sound sources with similar appearances,
(2) formulating a probabilistic framework including mixing levels as the latent variables to simultaneously estimate the number and locations of sound sources,
(3) deriving an objective function for integrating audio and visual information based on a spatial audio model, and
(4) experimentally confirming the applicability of the proposed method to real-world data recorded at a science museum using a robotic system.

\section{Related Work}

SSL has typically been formulated as an audio problem using a set of microphones, e.g., physical-model-based~\cite{music,phat,dsvd} and DNN-based methods~\cite{ssssl1,ssssl2}.
Integration of audio and visual information has also been studied to go beyond the audio-only settings, which is briefly reviewed in this section.
We also describe recent progress of self-supervised learning.

\subsection{Audio-Visual Sound Source Localization (AV-SSL)}

AV-SSL has been applied to various applications including speaker tracking~\cite{spk1,spk2}, traffic monitoring~\cite{car1}, and search-and-rescue tasks~\cite{rescue1}.
Since audio SSL and visual object detection have a similar purpose (i.e., finding the position of some objects), the standard strategy is to integrate methods for each task, which often requires a pre-trained DNN for visual object detection \cite{superb}.
Because of this requirement of the pre-trained DNN, existing literature mainly focuses on human \cite{spk1,spk2} or the objects whose well-established dataset of labeled images are relatively easy to obtain.
The focus of this paper is in a more general setting: the sounding objects to be detected include not only humans but also objects whose labeled dataset is difficult to obtain due to their excess variations.
Therefore, supervised training as in \cite{spk1,spk2} cannot be applied to our situation.

\subsection{Self-supervised Learning with Audio and Visual Data}

Self-supervised learning allows us to avoid the requirement of labels on the training dataset, which is quite desirable property for AV-SSL aiming at various types of objects.
Some early investigations have utilized monaural audio signals~\cite{ssavssl1,ssavssl2,ssavssl3} which do not contain any spatial information.
These methods are based on the contents in the observation, which requires differences in contents such as appearance and type of sound.
A recent study distinguishes sound sources from the temporal motion of corresponding objects (e.g., movements of violins)~\cite{ssavsslm}.
Our proposal aims at another direction, which distinguishes sound sources with similar appearance based on the spatial information (contained in a $360^\circ$ image and multichannel audio signal).

Some recent studies on AV-SSL utilize multichannel audio signals with self-supervised learning~\cite{ssavsslmulti2,ssavsslmulti1}.
These methods utilize teacher-student learning techniques, where a pre-trained visual DNN produces training targets of the audio DNN for localization.
That is, while an audio DNN does not need manual supervision, these self-supervised methods require labeled data for visual DNN.
In contrast, our method trains both audio and visual DNNs in a fully-self-supervised manner, i.e., no label is necessary for the training thanks to our probabilistic framework of the spatial audio model.

Meanwhile, spatial models have also been utilized for fully-self-supervised learning in each modality.
For monocular depth estimation, self-supervised learning is an active research topic because oracle depth maps are not easy to obtain~\cite{unsde2}.
Self-supervised learning of neural sound source separation has also been investigated by using a spatial model of multichannel audio signals~\cite{unsss1,unsss2}.
These methods do not require clean source signals, which are often unavailable in the real recordings.
They utilize a complex Gaussian mixture model (cGMM)~\cite{cgmm,bays1,bays2} and train DNNs to maximize the marginal likelihood or evidence lower bound (ELBO) of the probabilistic model.
Inspired from these existing works, our self-supervised method for AV-SSL aims to maximize the ELBO of the variant of a cGMM.
\section{Self-supervised Training of \\ Audio-Visual Sound Source Localization}

In order to achieve localization of objects with similar appearance, we present a self-supervised training method for multichannel neural AV-SSL that uses pairs of $360^\circ$ images and the corresponding multichannel audio mixtures.
The self-supervised training is formulated in a probabilistic manner based on a spatial audio model called a cGMM as in~\cite{unsss2}.

\subsection{Problem Specification}
\label{sec:problme}
\begin{figure*}[t]
    \centering
    \includegraphics[width=0.9\hsize]{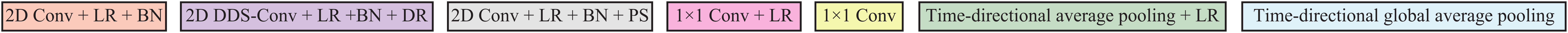}
    \vspace{-2.8mm} \\
    \subfloat[Visual DNN]{\includegraphics[width=0.48\hsize]{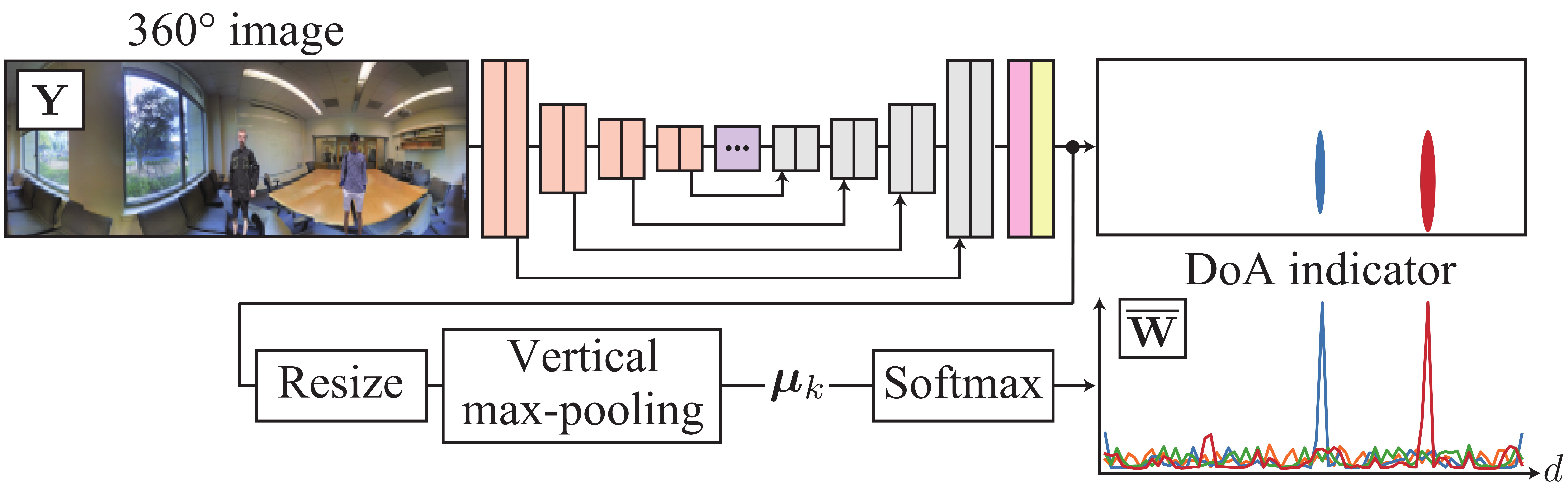}}
    \hfill
    \subfloat[Audio DNN]{\includegraphics[width=0.48\hsize]{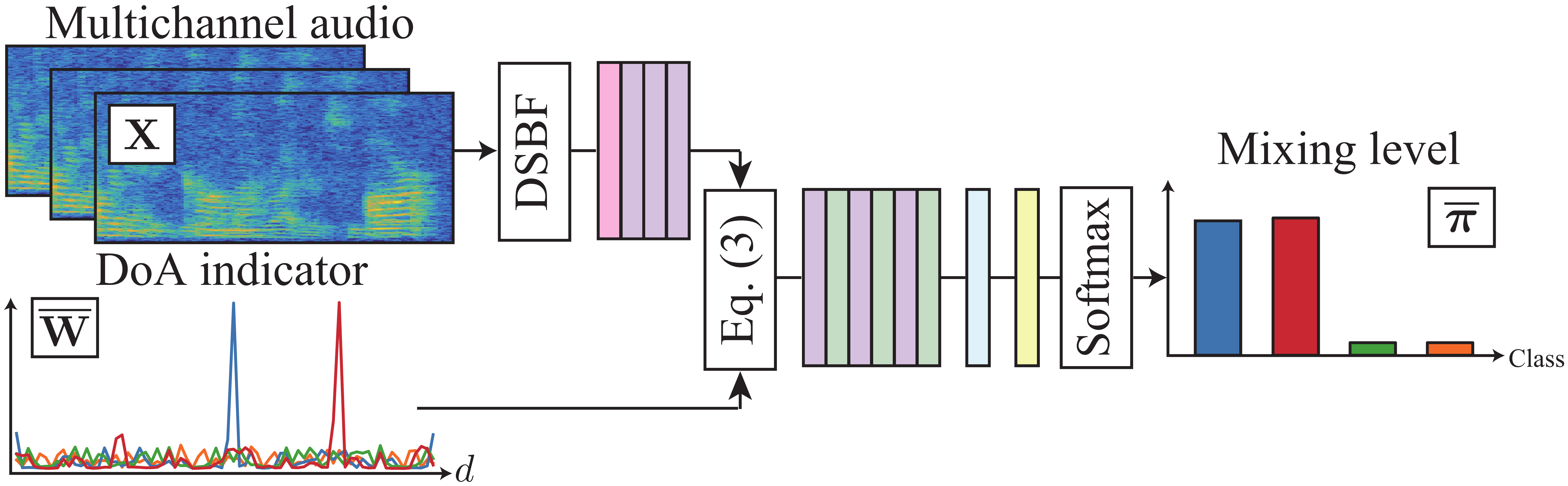}}
    \vspace{-2mm}
    \caption{Architectures of visual and audio DNNs.
    Here, LR, BN, DR, PS, and DDS-Conv indicate the leaky-ReLU, batch normalization, dropout, pixel shuffler, and dilated depthwise-separable convolution, respectively.
    }
    \label{fig:dnn}
    \vspace{-2.4mm}
\end{figure*}

The problem setting of our self-supervised training for multichannel neural AV-SSL is defined as follows: \\
\vspace{-6mm} \\
\rule{\hsize}{1pt}
{\bf Training data:} \\
$N$ pairs of (1) $360^\circ$ images $\y*^{(n)} \in \setRp^{I \times J \times 3}$ captured by a $360^\circ$ camera, and 
(2) multichannel audio mixtures $\x@^{(n)} \in \setC^M$ recorded by an $M$-channel microphone array. \\
{\bf Training targets:} \\
(1) visual DNN $\mathscr{G}$ that takes a $360^\circ$ image as input and estimates directions of sound source candidates, and \\
(2) audio DNN $\mathscr{H}$ that estimates mixing levels of sound source candidates from the estimated directions and multichannel audio mixture.  \\
{\bf Assumptions:} \\
(1) the configuration of the microphone array is given, \\
(2) movements of sources are negligibly small in $T$ frames. \\
\vspace{-6mm} \\
\rule{\hsize}{1pt}
where $i = 1, \ldots, I$ and  $j = 1, \ldots, J$ are respectively the vertical and horizontal indices for an image, and $t = 1, \ldots, T$ and $f = 1, \ldots, F$ are respectively the time and frequency indices for a multichannel audio spectrogram.
The suffix $(n)$ is hereinafter omitted because the objective function and the networks are independently defined for each of the $N$ data.
Under the first assumption, we can obtain the steering vectors $\svt@ \in \setC^M$ for potential directions $d = 1, \ldots, D$ and associate the directions of sounds and the locations of the corresponding objects.
We set $D$ potential directions at the same intervals on a horizontal plane for simplicity.
Under the second assumption, this paper treats not a video stream but an image.
The extension to handle a video stream is included in our future work.
Here, we allow that there exist objects that do not produce sounds (e.g., persons not speaking), which is not considered in conventional audio-visual self-supervised learning.
In addition, we also allow the number of sound sources is unknown.

\subsection{System Overview of Multichannel Neural AV-SSL}
\label{sec:net}

Our multichannel neural AV-SSL system is composed of two DNNs: visual DNN $\mathscr{G}$ and audio DNN $\mathscr{H}$ (Fig.~\ref{fig:dnn}).

\subsubsection{Visual DNN}

The visual DNN $\mathscr{G}$ estimates the DoA indicator $\overline{\mathbf{W}} = [\overline{\mathbf{w}}_1, \ldots, \overline{\mathbf{w}}_K]^\mathsf{T}$, where
$K$ is the number of potential sound sources, and $\overline{\mathbf{w}}_k = [\overline{w}_{k1}, \ldots, \overline{w}_{kD}]^\mathsf{T} \in \setRp^{D}$ indicates how the direction $d$ is dominated by source $k$:
\begin{align}
    \overline{\mathbf{w}}_k = \mathrm{softmax}(\bm{\mu}_k) = \mathrm{softmax}(\mathscr{G}(\mathbf{Y})_k),
    \label{eq:localization}
\end{align}
where $\boldsymbol{\mu}_k \!=\! [\mu_{k1}, \ldots, \mu_{kD}]^\mathsf{T}$ is the output of $\mathscr{G}$.
The main part of the visual DNN is similar to the U-Net architecture \cite{unet} that is followed by additional $1\times1$ convolutional layers, resize, and vertical max pooling as shown in Fig.~\ref{fig:dnn}-(a).
The output of the U-Net module represents activation maps of source candidates within an input image.
To associate the 2D activations to the 1D potential directions on a horizontal plane, the vertical max pooling is applied to the map.

\subsubsection{Audio DNN}

The audio DNN predicts the mixing levels $\overline{\zrate@} \in \setRp^{K}$ from the observed $M$-channel audio mixture $\mathbf{X} \in \setC^{T \times F \times M}$ and the predicted DoA indicators $\overline{\w*}$:
\begin{equation}
    \overline{\zrate@} = \mathscr{H}(\mathbf{X}, \overline{\w*}).
\end{equation}
In the audio DNN $\mathscr{H}$ (Fig.~\ref{fig:dnn}-(b)), we first apply delay-and-sum beamforming (DSBF) for every potential direction $d$ and extract their amplitude as $\mathbb{U}\bigl[\log(|\svt@^{\adj} \x@|)\bigr]$, where $\mathbb{U}$ is the frequency-wise mean and variance normalization.
The beamforming results are converted to direction-wise features $\mathbf{u}_{cd} \in \setR^{T}$ by convolutional block where the frequencies are treated as channels, and $c$ denotes the feature index.

To associate the audio and visual information, the audio features related to directions $\mathbf{u}_{cd}$ are transformed to those related to sources $\mathbf{v}_{ck} \in \setR^T$ by using DoA indicators $\overline{\w*}$:
\begin{equation}
    \mathbf{v}_{ck} = \sum_{d=1}^D \overline{w}_{kd} \mathbf{u}_{cd}.
    \label{eq:combert}
\end{equation}
The mixing levels $\zrate*$ are then obtained by passing the source-wise feature $\mathbf{v}_{ck}$ to another convolutional block, time-directional global average pooling, and affine block.

\subsubsection{Inference of Multichannel Neural AV-SSL System}
At the inference, our AV-SSL system predicts DoAs in the following two steps.
We first obtain DoA indicators $\overline{\w*}$ and mixing levels $\overline{\zrate*}$ by passing the observation through the visual and audio DNNs.
Then, we determine whether each source candidate $k$ actually produces sound by thresholding the DoA indicator and mixing level.
Finally, the DoA of each source $d_k^\star \in \{1,\ldots,D\}$ is calculated as follows:
\begin{equation}
    d_k^\star = \mathrm{arg}\max_d \,\,\,\overline{w}_{kd}.
\end{equation}

\subsection{Generative Model of Multichannel Audio Signal}
\label{sec:model}

To associate the spatial information in audio and visual observations, we formulate a cGMM-based spatial model of a multichannel mixture signal.
The cGMM has been known to robustly work in real-world environments~\cite{cgmm,chime3}.
As in the original cGMM, an observed signal $\x@ \in \setC^M$ is represented with $K$ source signals $\s \in \setC$ by
\begin{align}
    \x@ = \sum_{k=1}^{K} \z \cdot \sv@ \s, \label{eq:mixing}
\end{align}
where $\z \in \{0,1\}$ is a time-frequency (T-F) mask that satisfies $ \sum_{k=1}^{K} z_{tfk}=1$, and $\mathbf{a}_{fk} \in \setC^M$ is the steering vector of source $k$ at  frequency $f$.
The T-F mask $\z$ is introduced by assuming the source spectra sufficiently sparse in the T-F domain, and is assumed to follow a categorical distribution (denoted as $\mathrm{Cat}$):
\begin{align}
    \z@ = [\z[tf1], \ldots, \z[tfK]]^\T \sim  \distcategorical{\zrate[1], \ldots, \zrate[K]}, \label{eq:mask}
\end{align}
where $\pi_k \in \setRp$ is the mixing level of source $k$ that satisfies $\sum_k \pi_k = 1$.
Each source signal $\s$ is assumed to follow a complex Gaussian distribution characterized with a power spectral density $\psd \in \setRp$ as follows:
\begin{align}
    \s \sim \distcmpnormal{0}{\psd}.
\end{align}
Based on these assumptions, the observation $\x@$ follows a multivariate complex Gaussian mixture distribution given by
\begin{align}
    \x@ \sim \sum_{k=1}^{K} \zrate \distcmpnormal{\bm{0}}{\psd \scm}, \label{eq:multi}
\end{align}
where $\scm = \E[\sv@ \sv@^\adj ] \in \setC^{M \times M}$ is the spatial covariance matrix (SCM) of source $k$.
To deal with the unknown number of sources, we encourage the shrinkage of redundant source classes by putting a Dirichlet distribution (denoted as $\mathrm{Dir}$)~\cite{bays1} on the mixing levels:
\begin{align}
    [\zrate[1], \ldots, \zrate[K]]^\T \sim \distdirichlet{\alpha_0, \ldots, \alpha_0},
\end{align}
where $\alpha_0 \in \setRp$ is a hyperparameter.

To associate the multichannel audio observation $\x*$ and DoA candidates $\w*$ estimated by the visual DNN, we represent the SCM of source $k$ by the weighted sum of template SCMs $\scmt = \svt@ \svt@^\adj$ for potential directions $d$ as follows:
\begin{align}
    \scm \approx \sum_{d=1}^{D} \w \scmt + \epsilon\eye, \label{eq:steering}
\end{align}
where $\epsilon$ is a small number to ensure $\scm$ positive definite.
To prevent the estimates of $\w$ from taking exceedingly large values, we put the following log-normal prior on $\w$: 
\begin{align}
    \w \sim \distlognormal{0}{\sigma_0^2}, \label{eq:prioriw}
\end{align}
where $\sigma_0 \in \setRp$ is a scale parameter.

\subsection{Self-supervised Training Based on Variational Inference}
\label{sec:training}
\begin{figure}[tbp]
  \centering
  \vspace{1mm}
  \includegraphics[width=0.95\hsize]{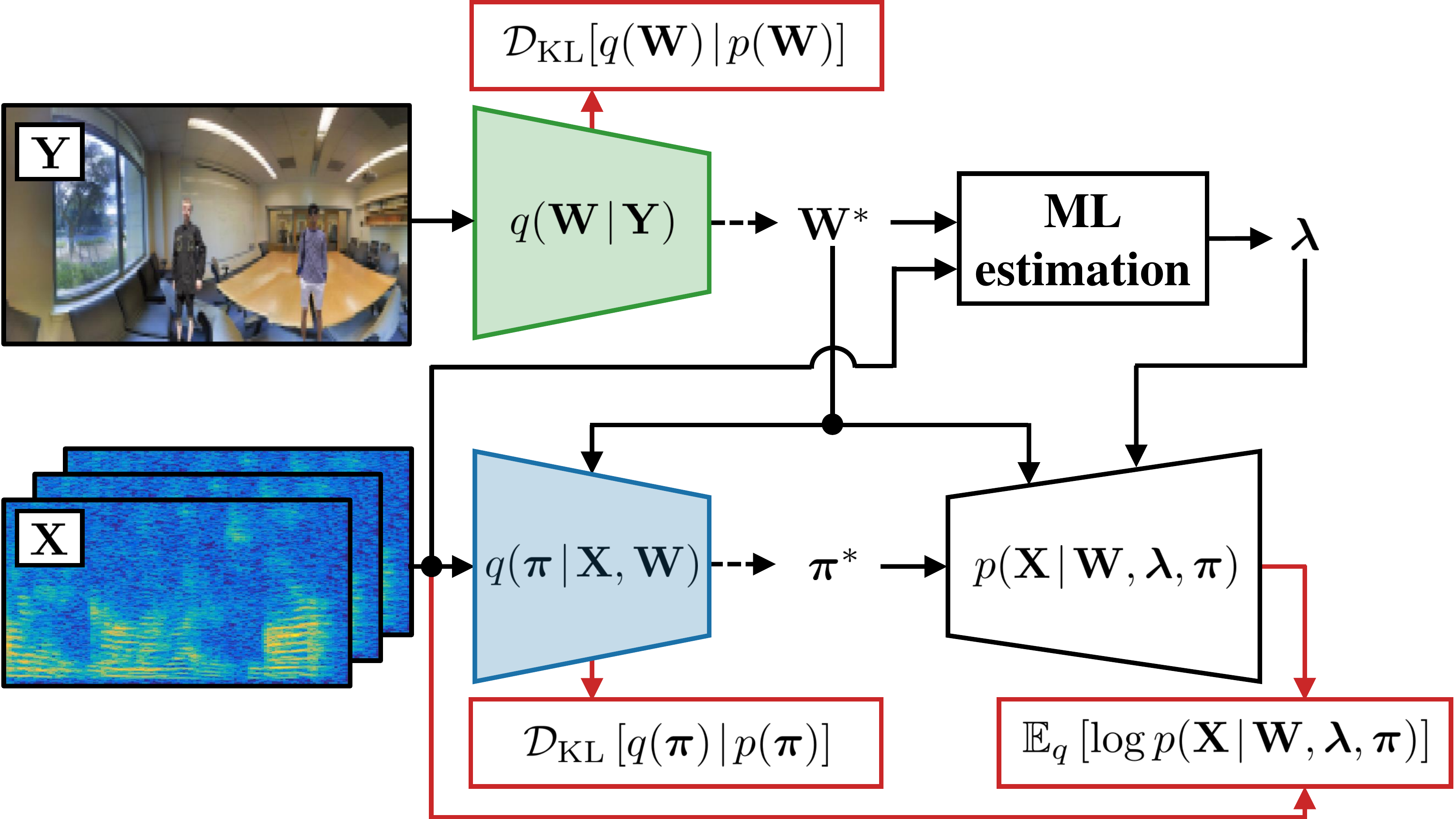}
  \caption{Overview of our self-supervised training based on probabilistic spatial audio model. The green and blue blocks are the visual and audio DNNs, respectively.}
  \label{fig:overview}
\end{figure}

The audio and visual DNNs are trained in a self-supervised manner so that they estimate the posterior distributions of $\zrate@$ and $\w*$, respectively (Fig.~\ref{fig:overview}).
We employ a Bayesian approach to treat the unknown number of sources in a unified framework~\cite{bays1}.
Specifically, we represent the posterior distributions with the outputs of the DNNs, $\boldsymbol{\mu}_k =\mathscr{G}(\mathbf{Y})_k$ and $\overline{\zrate@} = \mathscr{H}(\mathbf{X}, \w*)$, as follows:
\begin{align}
    q(\mathbf{W} \mid \mathbf{Y}) &= \prod_{k=1}^K\prod_{d=1}^D \distlognormal{\mu_{kd}}{\sigma^2_k}, \\
    q(\zrate@ \mid \mathbf{X}, \mathbf{W})&= \mathrm{Dir}(\beta\overline{\pi}_1, \ldots, \beta\overline{\pi}_K),
\end{align}
where $\sigma_k \in \setRp$ and $\beta \in \setRp$ are scale parameters representing the uncertainty and jointly optimized in this training.
We train the two DNNs so that the variationl posterior $q(\zrate@ | \mathbf{X}, \mathbf{W})q(\mathbf{W} | \mathbf{Y})$ approximates the true posterior $p(\mathbf{W}, \zrate@ | \mathbf{X}, \psd*)$
    by minimizing the Kullback--Leibler (KL) divergence between them.
This minimization, which is indeed intractable, is conducted by maximizing the ELBO~\cite{bays1} defined as follows:
\begin{align}
    \hspace{-2mm}\elbo = \E_q \left[ \log p(\x* \!\mid\! \w*, \!\psd*, \!\zrate*) \right] \!-\! \KL\!\left[ q(\w*, \!\zrate*) \!\mid\! p(\w*, \!\zrate*) \right].\!\!
    \label{eq:elbo}
\end{align}

Since the first term of Eq.~\eqref{eq:elbo} is difficult to calculate analytically, we utilize Monte Carlo approximation whose gradient can be obtained by using reparameterization trick~\cite{rep2}:
\begin{align}
&\E_q \left[ \log p(\x* \!\mid\! \w*, \!\psd*, \!\zrate*) \right] \approx \nonumber \\
&\hspace{5pt} \sum_{t=1}^T\sum_{f=1}^F \log \sum_{k=1}^K \frac{\ezrate}{|\psd \escm|} \exp \left( - \frac{\x@^\adj \escm^{-1} \x@ }{\psd} \right)
\label{eq:expect}
\end{align}
where $\escm = \sum_{d=1}^D\ew \scmt$ is a sampled SCM, and $\ew$ and $\ezrate$ are samples of $\ew \sim q(\w \mid \mathbf{Y})$ and $\ezrate \sim q(\zrate \mid \x*, \ew*)$, respectively.
The power spectral density $\lambda_{tfk}$ is substituted with a maximum likelihood estimate:
\begin{equation}
    \psd = \frac{1}{M} \x@^\adj \escm^{-1} \x@. \label{eq:lambda}
\end{equation}

The second term of Eq.~\eqref{eq:elbo} is calculated separately as
\begin{align}
    \KL\left[ q(\w*, \!\zrate*) \!\mid\! p(\w*, \!\zrate*) \right] &= \KL\left[ q(\w*) \mid p(\w*) \right]\nonumber \\
    &\hspace{-30pt}+ \mathbb{E}_{q(\w*)}\left[\KL\left[ q(\zrate*|\w*) \mid p(\zrate*) \right]\right]. \label{eq:klterm}
\end{align}
The first term in Eq.~\eqref{eq:klterm} is calculated analytically as follows:
\begin{align}
\!\!\! \KL \left[ q(\w*) \!\mid\! p(\w*) \right] \! = \! \sum_{k=1}^K \sum_{d=1}^D \frac{ \mu_{kd}^2 + \sigma_k^2 - \sigma_0^2}{2 \sigma_0^2} \!+\! \log\frac{\sigma_0}{\sigma_k}, \!\!
\label{eq:kllog}
\end{align}
where $\mu_{kd}$ is the output of the visual DNN $\mathscr{G}$.
The second term is approximately calculated by using the Monte Carlo sampling as follows:
\begin{align}
        \!\!\!&\mathbb{E}_{q(\w*)}\left[ \KL \left[ q(\zrate* \!\mid\! \w*) \!\mid\! p(\zrate*) \right]\right] \approx \log\frac{\Gamma(\beta\overline{\pi}_\cdot)}{\Gamma(K\alpha_0)} \nonumber \\[-5pt]
        \!\!\!& -\!\! \sum_{k=1}^K \log\frac{\Gamma(\beta\overline{\pi}_k)}{\Gamma(\alpha_0)} \!+\!\!\sum_{k=1}^K (\beta\overline{\pi}_k \!-\! \alpha_0) \!\left\{\psi(\beta\overline{\pi}_k) \!-\! \psi(\beta\overline{\pi}_\cdot)\right\}\!,\!\!\!
\label{eq:kldir}
\end{align}
where $\overline{\zrate*}$ is calculated with the sampled DoA indicators $\w*^*$,
$\overline{\pi}_{\cdot}$ represents $\sum_{k=1}^K \overline{\pi}_k$, $\Gamma(\cdot)$ is the gamma function, and $\psi(\cdot)$ is the digamma function.

Based on Eqs.~\eqref{eq:expect}--\eqref{eq:kldir}, we can approximately calculate the ELBO $\elbo$ and train the audio and visual DNNs so that the ELBO is maximized.
This maximization can be conducted with stochastic gradient descent because all the above equations are differentiable.
Note that this training does not use any labels or oracle signals and is conducted in a fully-self-supervised manner as a Bayesian inference.
\section{Experimental Evaluation \\ with Simulated Data} \label{sec:evaluation}

We validated our self-supervised training method for multichannel neural AV-SSL with simulated indoor environments where multiple persons speak simultaneously.

\subsection{Experimental Configuration}

Inspired by a synthesis method~\cite{cap}, we generated pairs of $360^\circ$ images and corresponding multichannel audio mixtures.
We synthesized the images by utilizing indoor images from 2D-3D-S dataset~\cite{2d3ds} and person images from the Clothing Co-Parsing dataset~\cite{ccp}.
More specifically, persons were located and rendered $0.5$ m to $2.0$ m from the camera at random.
The multichannel audio signals, on the other hand, were generated by convoluting room impulse responses (RIRs) to monaural speech signals selected from the WSJ0 corpus~\cite{wsj0}.
Speech signals were cut to a $1.0$-second clip randomly, and mixed at random powers uniformly chosen between $-2.5$ dB and $2.5$ dB from a reference value.
The RIR was generated by using the image method~\cite{rir} where the reverberation time (RT$_{60}$) was chosen at random between $0.2$ s and $0.4$ s, and the room dimension was fixed to $5.0$ m $\times$ $5.0$ m $\times$ $3.0$ m for simplicity.
A circular microphone array with a diameter of $20$ cm was located at the center of the room.
Gaussian noise was added to the mixtures with the signal-to-noise ratio of $20$ dB to imitate diffuse noise.

To confirm the effectiveness of the audio DNN, we generated two types of datasets.
(1) The first one contains two or three persons in each image, and all persons speak.
(2) The other one contains two to four persons, but only two or three speak for simulating the situation where several persons were actually not speaking.
For each of dataset types, we generated three datasets with the diffetent number of microphones $M \in \{2, 4, 6\}$.
For all conditions, we generated $20000$ pairs of simulated data for training and $1000$ pairs for testing, respectively.
The images were downsampled to $88\times288$ and the audio signals were sampled at $16$-kHz.

\begin{table}[t]
    \centering
    \vspace{1mm}
    \caption{Localization performance in F-measure under the first condition in which all persons spoke.}
    \label{tab: fval1}
    \vspace{-1.5mm}
    
    \setlength{\tabcolsep}{5pt}
    \begin{tabular}{c|cc|cc|cc}
        \toprule
        \# of mics. & \multicolumn{2}{c|}{$M = 2$} & \multicolumn{2}{c|}{$M = 4$} & \multicolumn{2}{c}{$M = 6$} \\
        \# of srcs. &\!$L = 2$\!&\!$L = 3$\!&\!$L = 2$\!&\!$L = 3$\!&\!$L = 2$\!&\!$L = 3$\! \\
        \midrule
        MUSIC & $ - $ & $ - $ & $0.80$ & $0.70$ & $0.84$ & $0.78$ \\
        SRP-PHAT & $0.41$ & $0.37$ & $0.65$ & $0.56$ & $0.74$ & $0.63$ \\
        \midrule
        w/o Aud & $0.68$ & $\bf 0.68$ & $0.85$ & $0.82$ & $\bf 0.85$ & $0.82$ \\
        Proposed & $\bf 0.74$ & $0.66$ & $\bf 0.86$ & $\bf 0.83$ & $\bf 0.85$ & $\bf 0.84$ \\
        \bottomrule
    \end{tabular}
\vspace{0.5mm}
    \centering
    \setlength\belowcaptionskip{1mm}
    \caption{Localization performance in F-measure under the second condition in which not all persons spoke.}
    \label{tab: fval2}
    \vspace{-1.5mm}
    
    \setlength{\tabcolsep}{5pt}
    \begin{tabular}{c|cc|cc|cc}
        \toprule
        \# of mics. & \multicolumn{2}{c|}{$M = 2$} & \multicolumn{2}{c|}{$M = 4$} & \multicolumn{2}{c}{$M = 6$} \\
        \# of srcs. &\!$L = 2$\!&\!$L = 3$\!&\!$L = 2$\!&\!$L = 3$\!&\!$L = 2$\!&\!$L = 3$\! \\
        \midrule
        MUSIC & $ - $ & $ - $ & $0.73$ & $0.64$ & $\bf 0.85$ & $\bf 0.77$ \\
        SRP-PHAT & $0.43$ & $0.36$ & $0.66$ & $0.54$ & $0.73$ & $0.63$ \\
        \midrule
        w/o Aud & $0.55$ & $0.54$ & $0.62$ & $0.63$ & $0.69$ & $0.67$ \\
        Proposed & $\bf 0.70$ & $\bf 0.70$ & $\bf 0.77$ & $\bf 0.73$ & $ 0.78$ & $ 0.74$ \\
        \bottomrule
    \end{tabular}
    \vspace{-2mm}
\end{table}

The visual and audio DNNs were jointly trained by using the AdamW optimizer for $200$ epochs with the learning rate of $1.0\times 10^{-3}$.
The hyperparameters of the cGMM $K$, $\alpha_0$, and $\sigma_0$ were set to $4$, $0.01$ and $1.0$, respectively.
The multichannel spectrograms were calculated by the short-time Fourier transform with a Hann window whose length of $512$ samples and time-shift of $160$ samples.
The number of potential directions was set to $D=72$, and the steering vector for each direction $\svt@$ was theoretically calculated under the plane-wave assumption.
The thresholding parameter for $\overline{\pi}_k$ was set to $0.02$.
The source classes corresponding to the diffuse noise were omitted by thresholding $\overline{w}_{kd}$. 
The threshold was determined such that the F-measure~\cite{fval} of localization results was maximized.
The architectures of the audio and visual DNNs are summarized in Fig.~\ref{fig:dnn}.
These parameters were determined experimentally.

The proposed SSL was compared with two existing audio-only methods.
One was the subspace-based method called multiple signal classification (MUSIC)~\cite{music}, and the other was steered-response power phase transform (SRP-PHAT)~\cite{phat}.
Although these methods were proposed more than 10 years ago, they are still being used in the state-of-the-art systems~\cite{locata}.
For MUSIC, the parameter corresponding to the number of sound sources was set to $2$ for $M=4$ and $3$ for $M=6$.
To validate the effectiveness of the audio DNN, we also evaluated a simplified version of our system in which the mixing level $\zrate$ was fixed to $1/K$, which is abbreviated here as w/o Aud.

\begin{table}[t]
    \vspace{1mm}
    \centering
    \caption{Sound source number estimation performance in correct rate under the second condition.}
    \label{tab:acc}
    \vspace{-1.5mm}
    
    \setlength{\tabcolsep}{5pt}
    \begin{tabular}{c|cc|cc|cc}
        \toprule
        \# of mics. & \multicolumn{2}{c|}{$M = 2$} & \multicolumn{2}{c|}{$M = 4$} & \multicolumn{2}{c}{$M = 6$} \\
        \# of srcs. &\!$L = 2$\!&\!$L = 3$\!&\!$L = 2$\!&\!$L = 3$\!&\!$L = 2$\!&\!$L = 3$\! \\
        \midrule
        MUSIC & $ - $ & $ - $ & $0.59$ & $0.27$ & $\bf 0.83$ & $\bf 0.59$ \\
        SRP-PHAT & $0.48$ & $0.01$ & $0.50$ & $0.17$ & $0.51$ & $0.26$ \\
        \midrule
        w/o Aud & $0.37$ & $0.42$ & $0.54$ & $0.42$ & $0.56$ & $0.51$ \\
        Proposed & $\bf 0.52$ & $\bf 0.58$ & $\bf 0.72$ & $\bf 0.45$ & $ 0.71$ & $0.47$ \\
        \bottomrule
    \end{tabular}
\end{table}

\subsection{Experimental Results}
\begin{figure*}[tbp]
  \vspace{2mm}
  \setlength\abovecaptionskip{1mm}
  \centering
  \includegraphics[width=0.32\hsize]{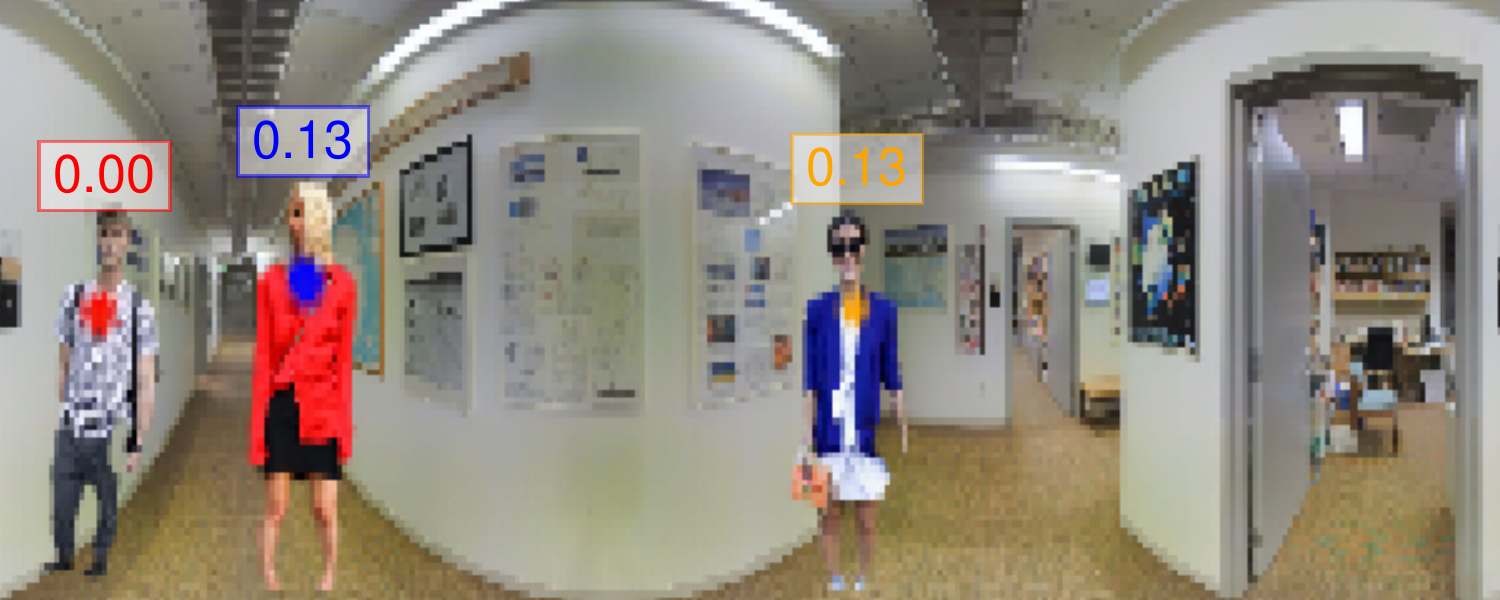} \hfill
  \includegraphics[width=0.32\hsize]{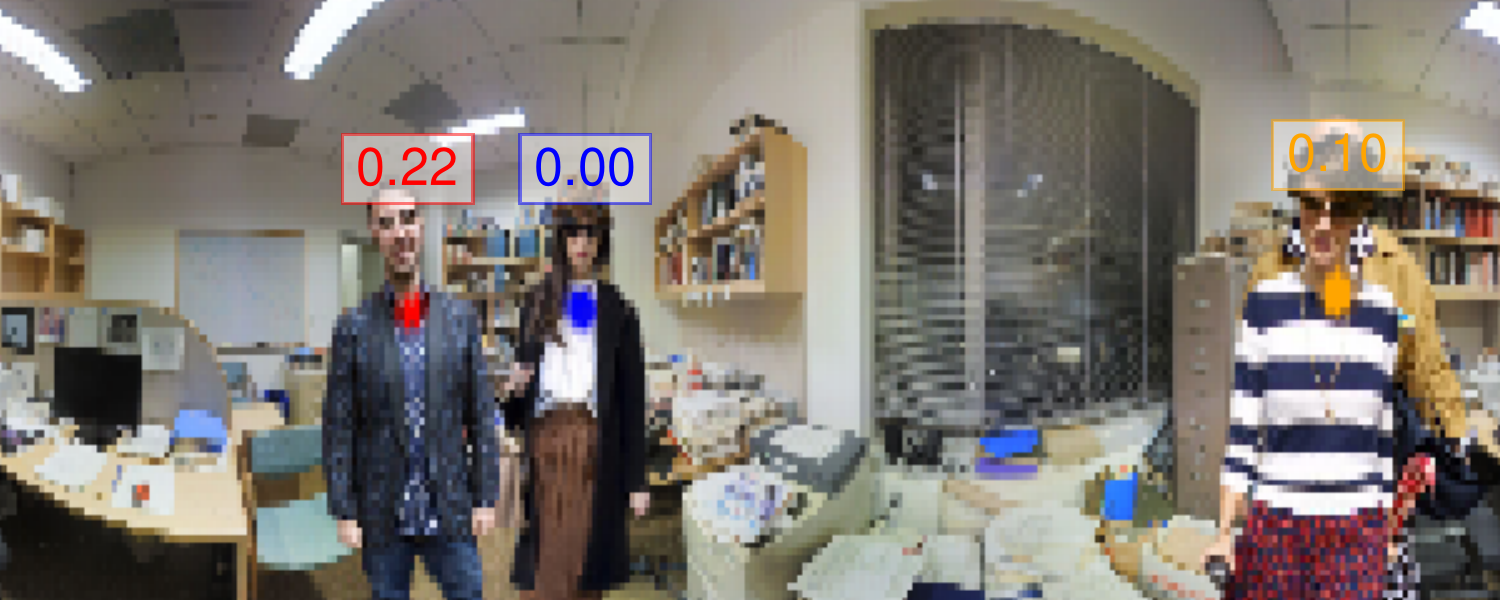} \hfill
  \includegraphics[width=0.32\hsize]{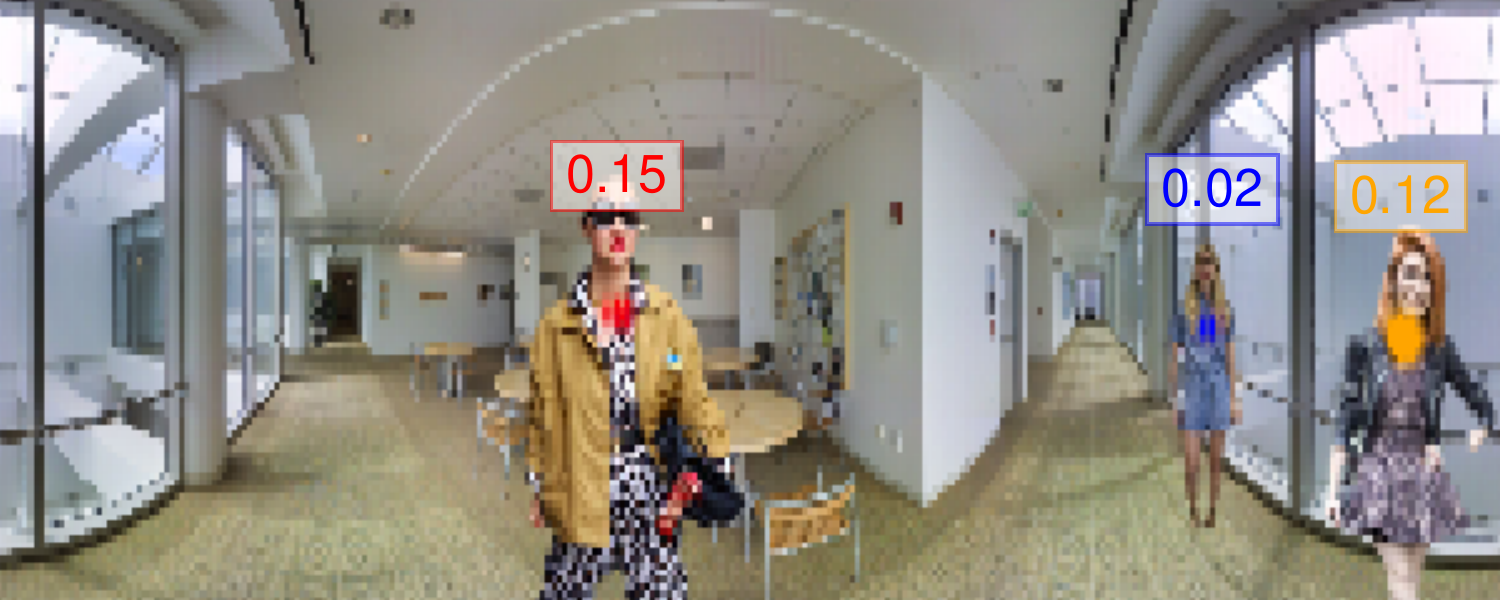}
  \\ \vspace{3mm}
  \includegraphics[width=0.32\hsize]{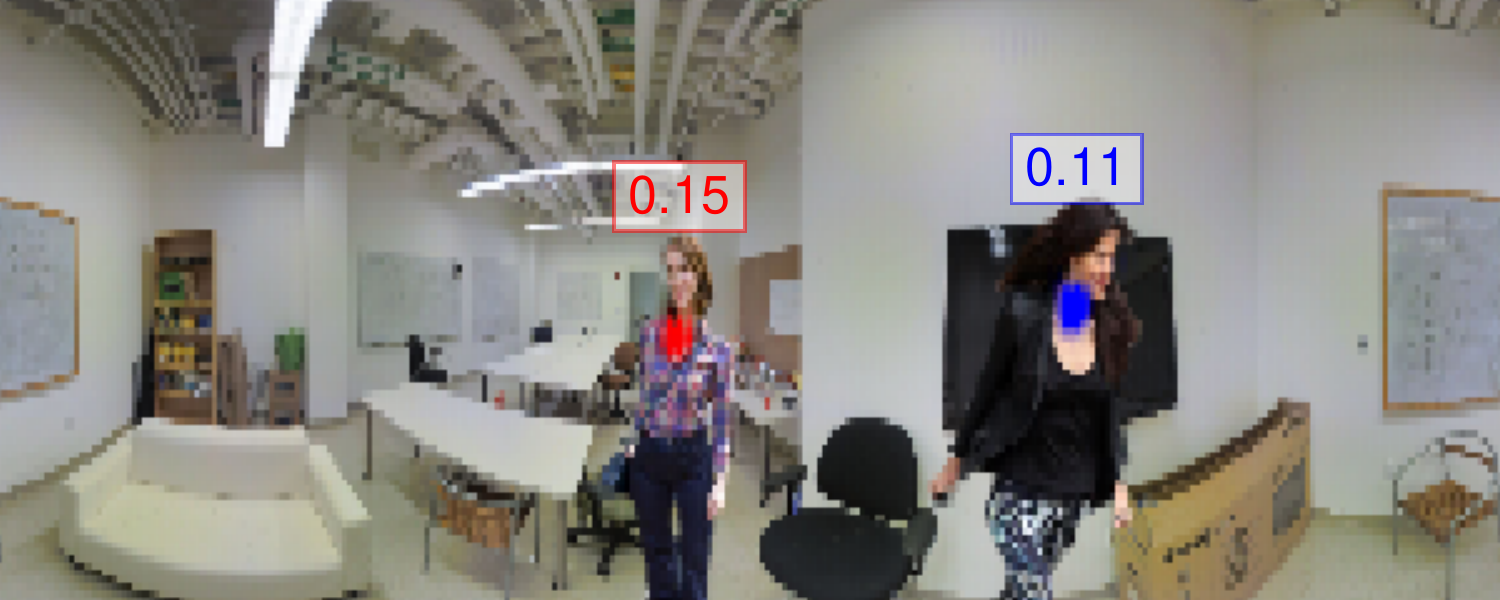} \hfill
  \includegraphics[width=0.32\hsize]{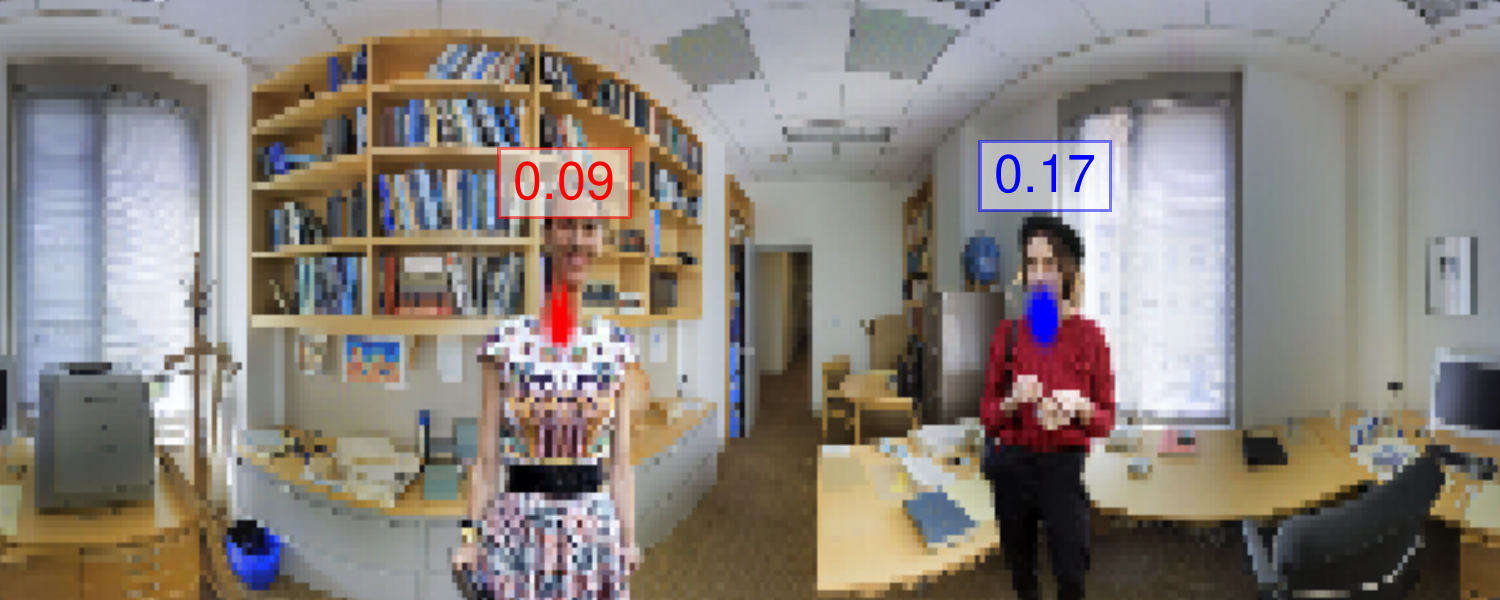} \hfill
  \includegraphics[width=0.32\hsize]{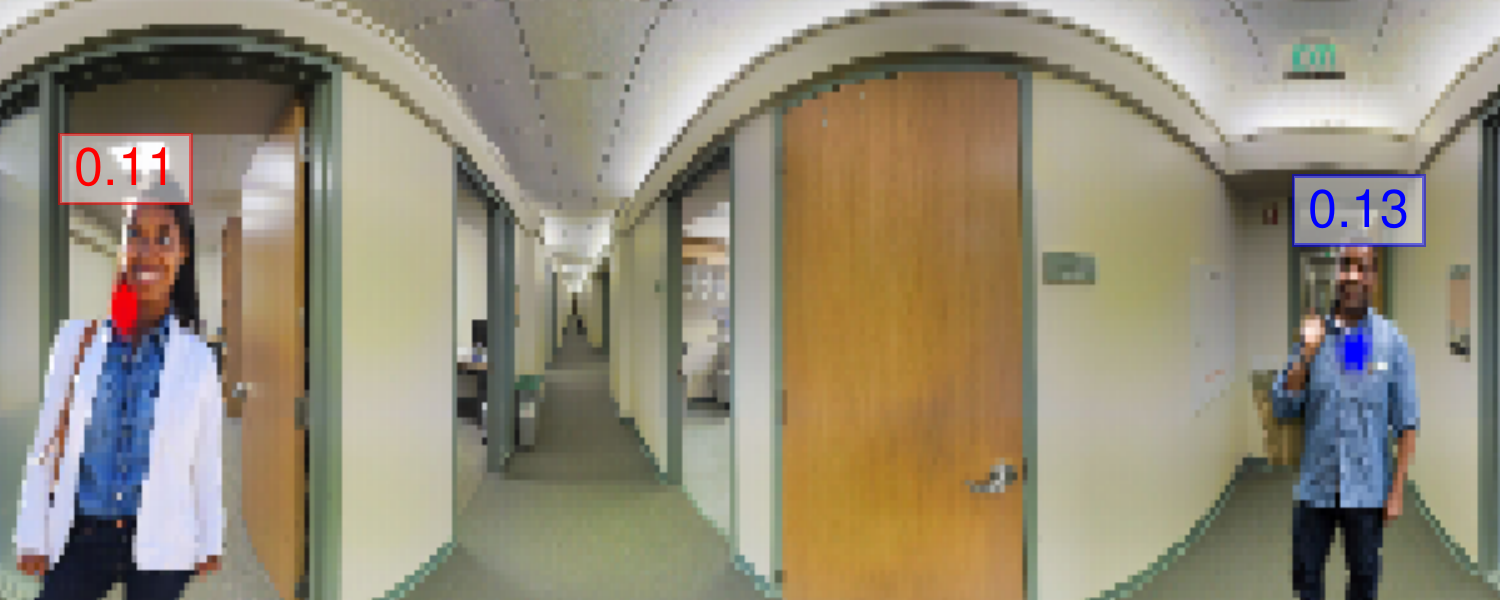}
  \caption{Visualization of activation maps obtained by visual DNN from our simulated indoor dataset.
  Each color corresponds to each sound source candidate.
  Estimated mixing levels are shown in the boxes.}
  \label{fig:focus}
  \vspace{-1.5mm}
\end{figure*}

Table~\ref{tab: fval1} shows the SSL performance in F-measure at the first condition in which all persons speak.
Here, $L$ is the number of actual sound sources.
Our multichannel neural AV-SSL system was comparable to the conventional methods when $M=6$.
When the number of microphones decreases, the conventional methods were significantly degraded. 
On the other hand, our AV-SSL system retained its performance thanks to the visual information.
In this condition, the proposed method and that without the audio DNN achieved similar performance except $M=2$.

The SSL performance under the second condition is shown in Table~\ref{tab: fval2}.
Under this condition, our AV-SSL system should determine whether or not each person actually spoke.
The performance of our system without the audio DNN was significantly decreased because it cannot distinguish persons who did not speak.
On the other hand, the proposed system with the audio DNN still performed well.
We also evaluated the correct rate of the source number estimation under this condition.
The results are shown in Table~\ref{tab:acc}.
The proposed system with the audio DNN outperformed that without the audio DNN except $L=3$ with $M=6$.
These results show the importance of the audio DNN that determines whether or not each sound source candidate produces sound.

To confirm that the visual DNN detects the sound source objects properly, we visualized the activation map of the visual DNN (Fig.~\ref{fig:dnn}-(a)) under the second condition with $M=6$.
The results are shown in Fig.~\ref{fig:focus}.
Each colored area corresponds to each estimated sound source candidate, and the estimated mixing levels $\overline{\zrate*}$ were shown in boxes.
We can see that the DNN distinguished multiple persons properly with $L=\{2,3\}$.
In addition, we confirmed that the persons whose estimated mixing levels were $0.00$ did not produce sounds in the left and center on the first row of Fig.~\ref{fig:focus}.

\section{Real-data Analysis \\ with Autonomous Robot in Science Museum}
We demonstrate our self-supervised method with real data recorded by an autonomous robot called Peacock (Fig.~\ref{fig:peacock}).

\subsection{Experimental Setup}

 \begin{figure}
    \setlength\abovecaptionskip{2mm}
    \centering
    \includegraphics[width=0.95\hsize]{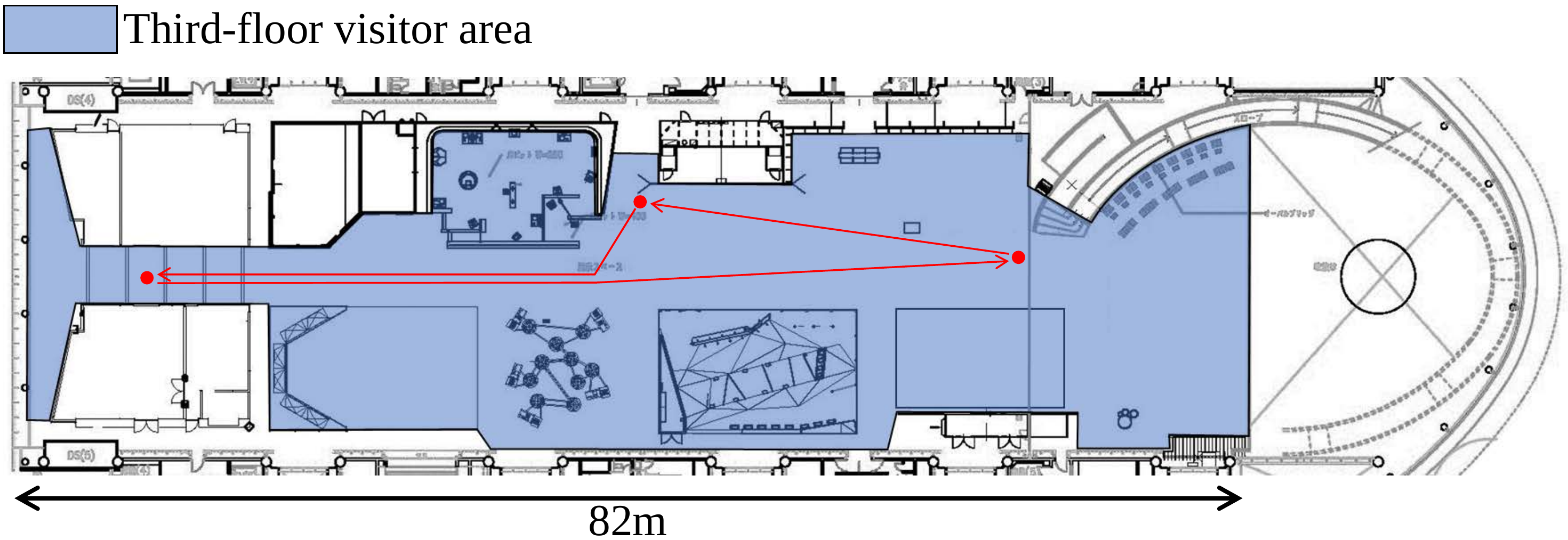}
    \caption{Third floor of Miraikan. Peacock moved along red arrows while avoiding visitors.}
    \label{fig:miraikan}
\end{figure}
\begin{figure*}[t]
    \vspace{2mm}
    \centering
    \includegraphics[width=0.325\hsize]{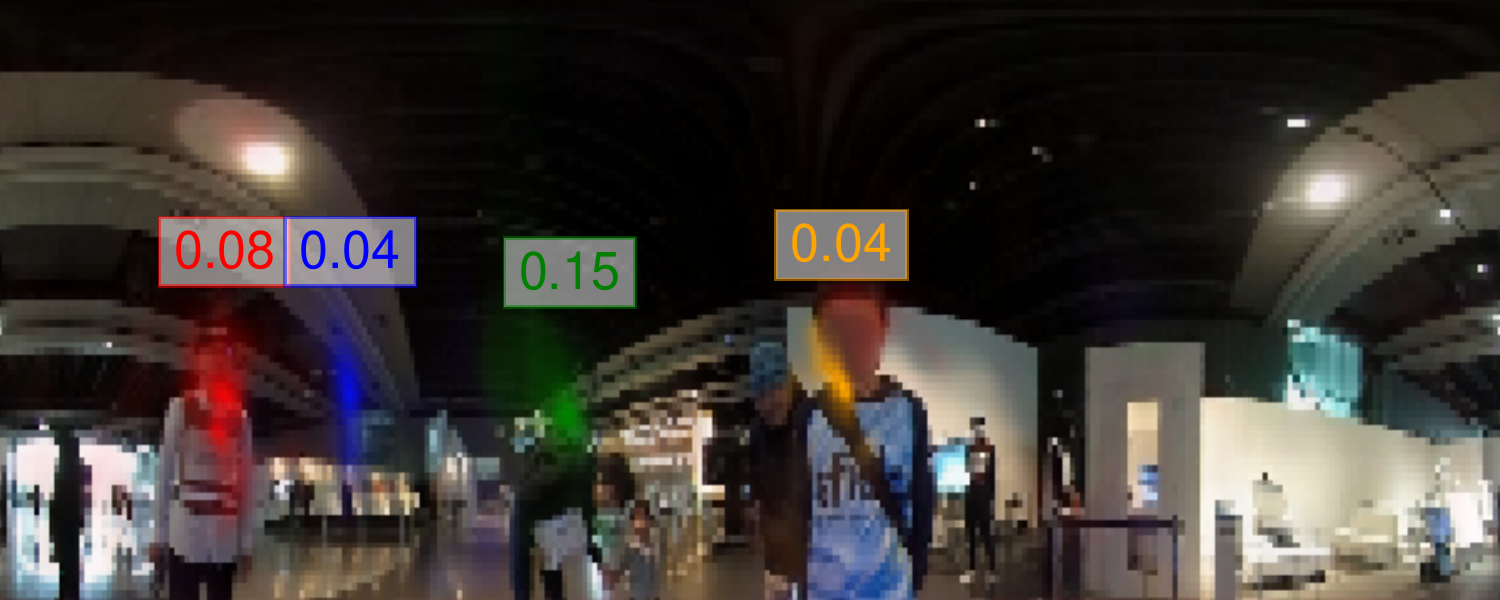} \hfill
    \includegraphics[width=0.325\hsize]{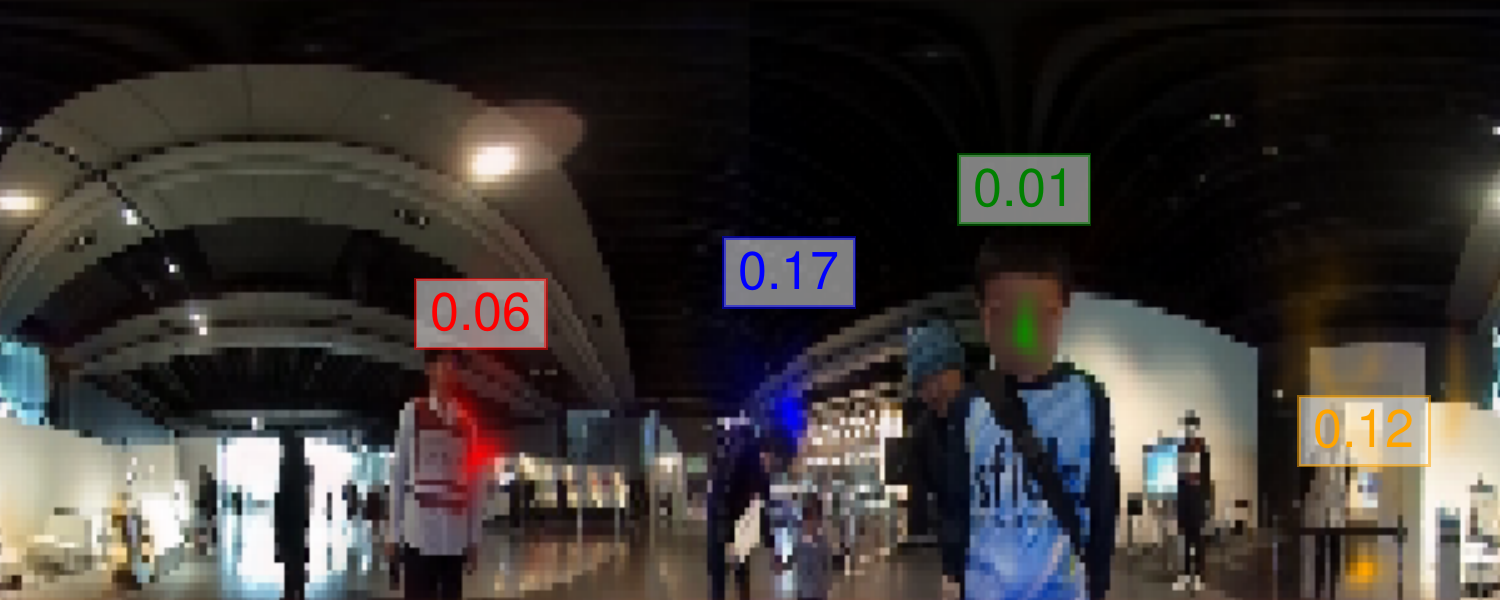} \hfill
    \includegraphics[width=0.325\hsize]{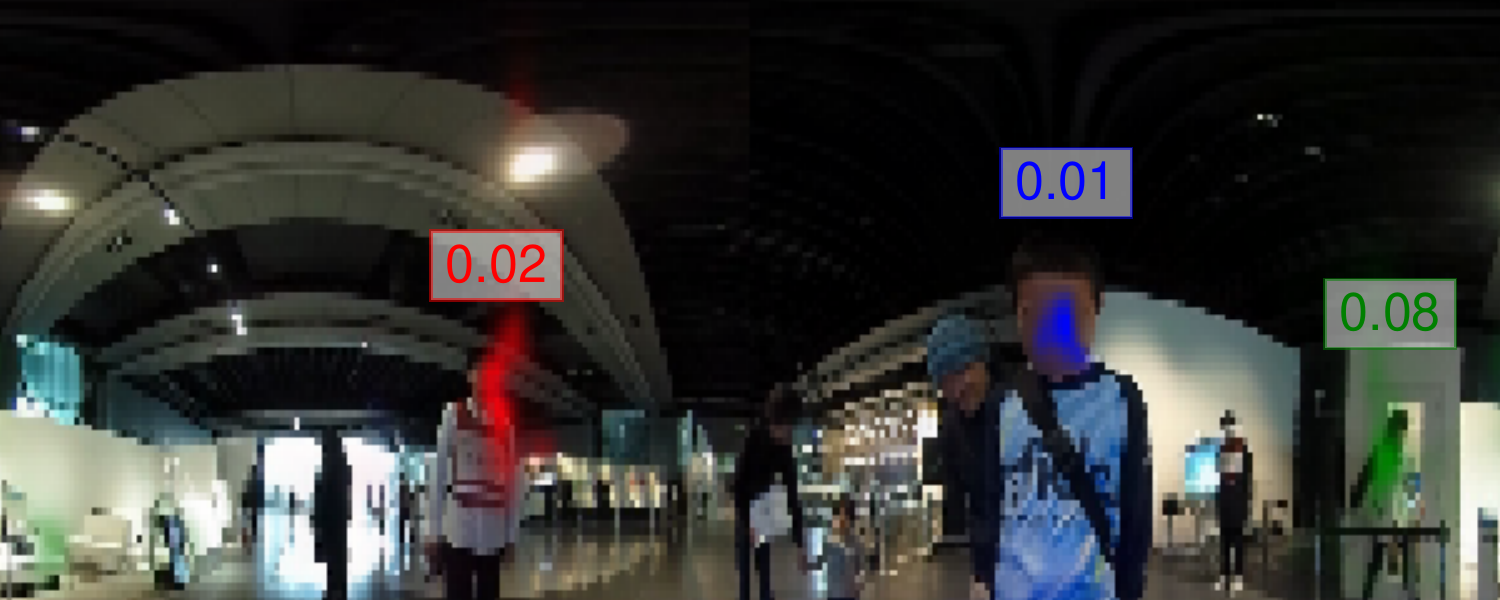} \\[1.5mm]
    \includegraphics[width=0.325\hsize]{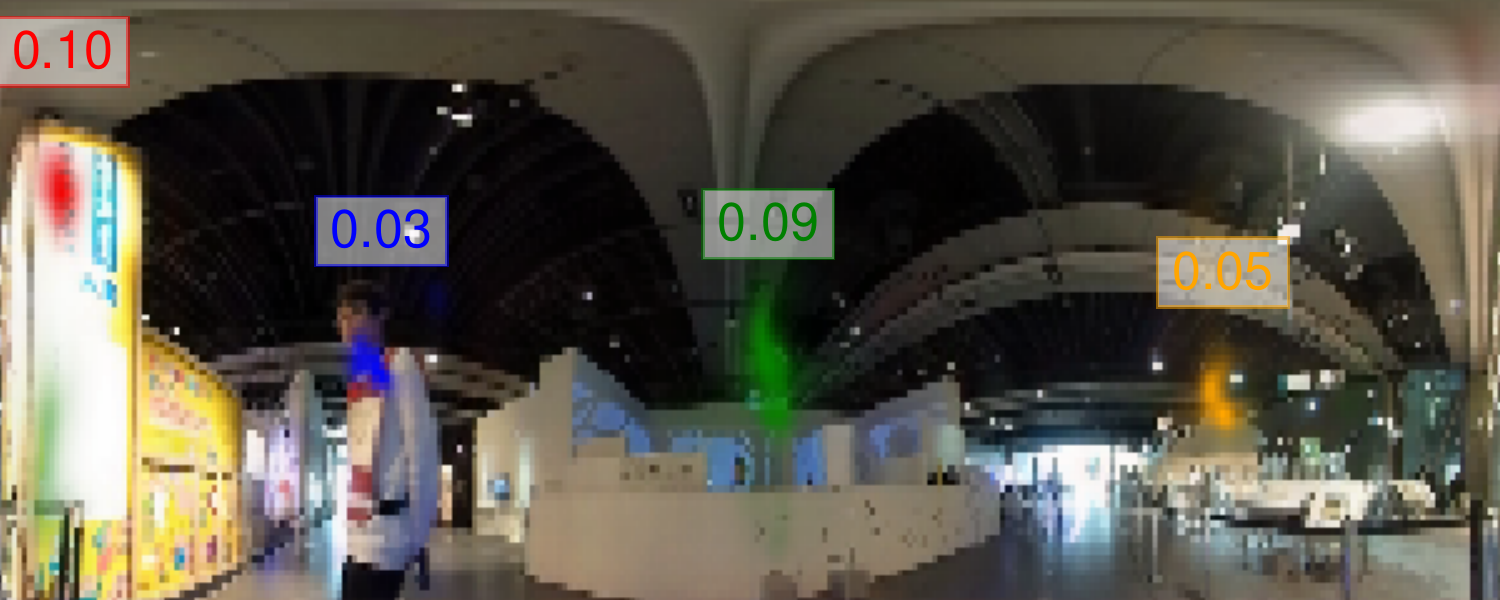} \hfill
    \includegraphics[width=0.325\hsize]{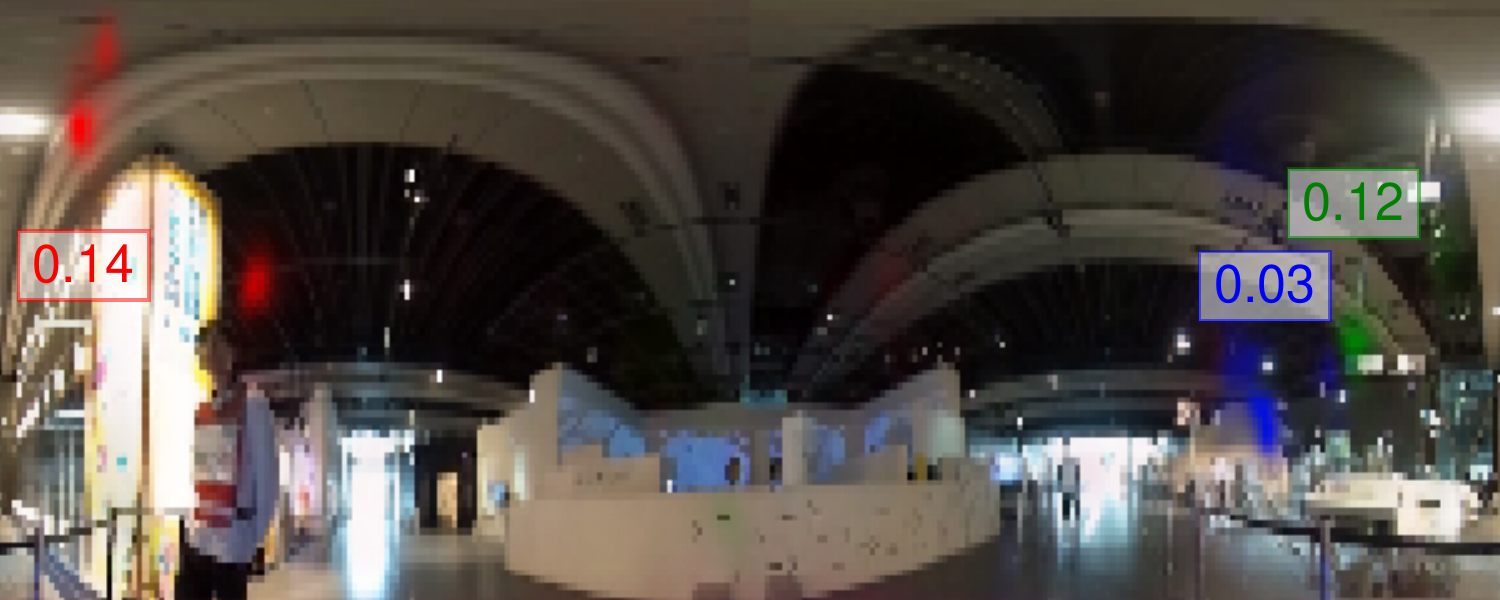} \hfill
    \includegraphics[width=0.325\hsize]{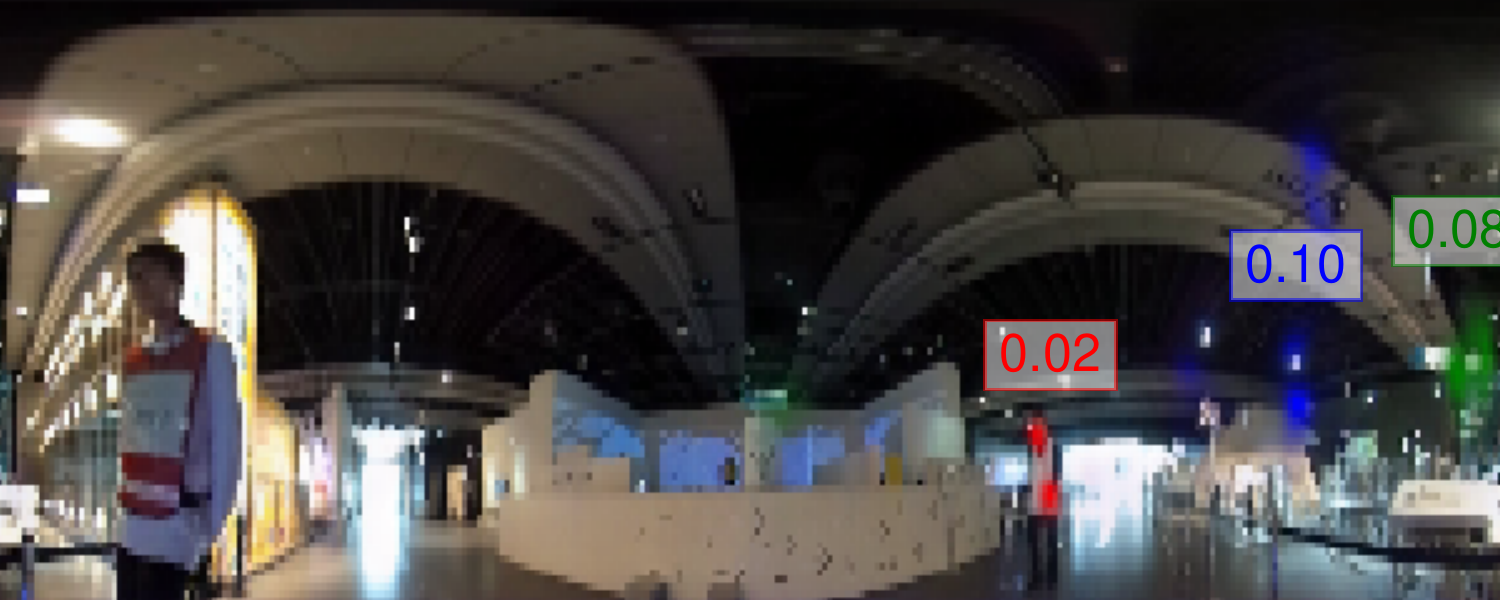} \\[1.5mm]
    \includegraphics[width=0.325\hsize]{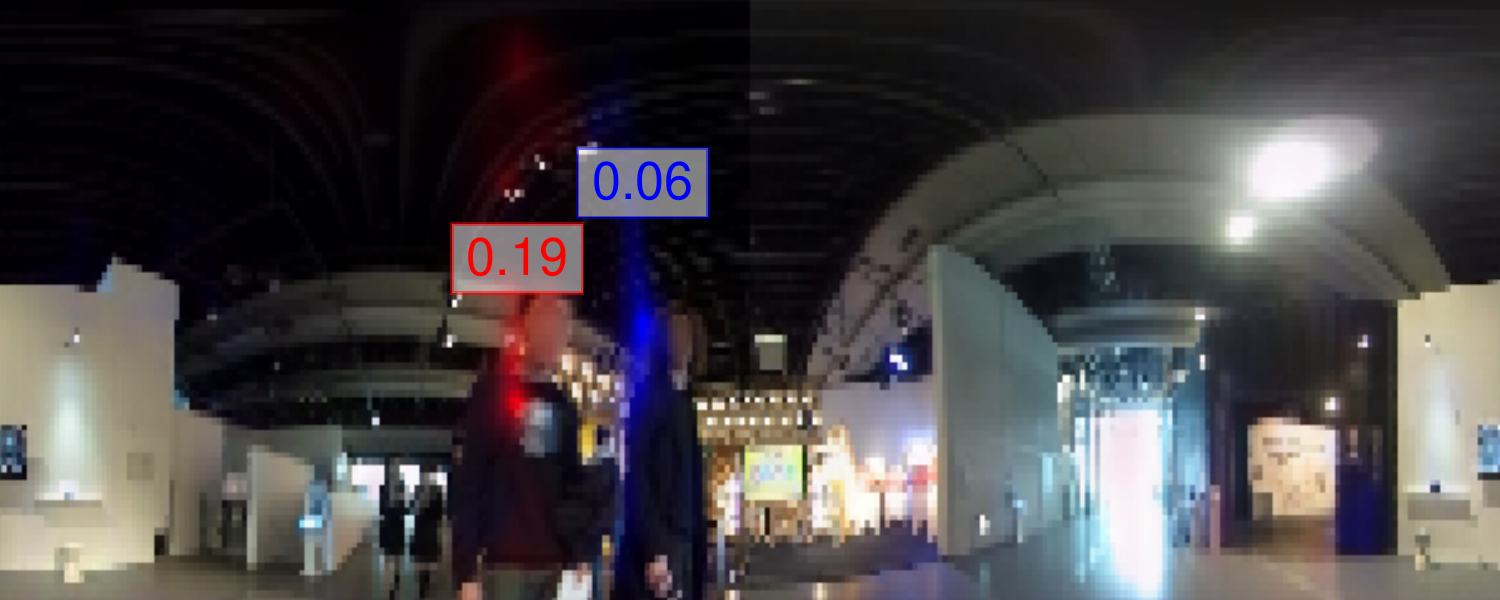} \hfill
    \includegraphics[width=0.325\hsize]{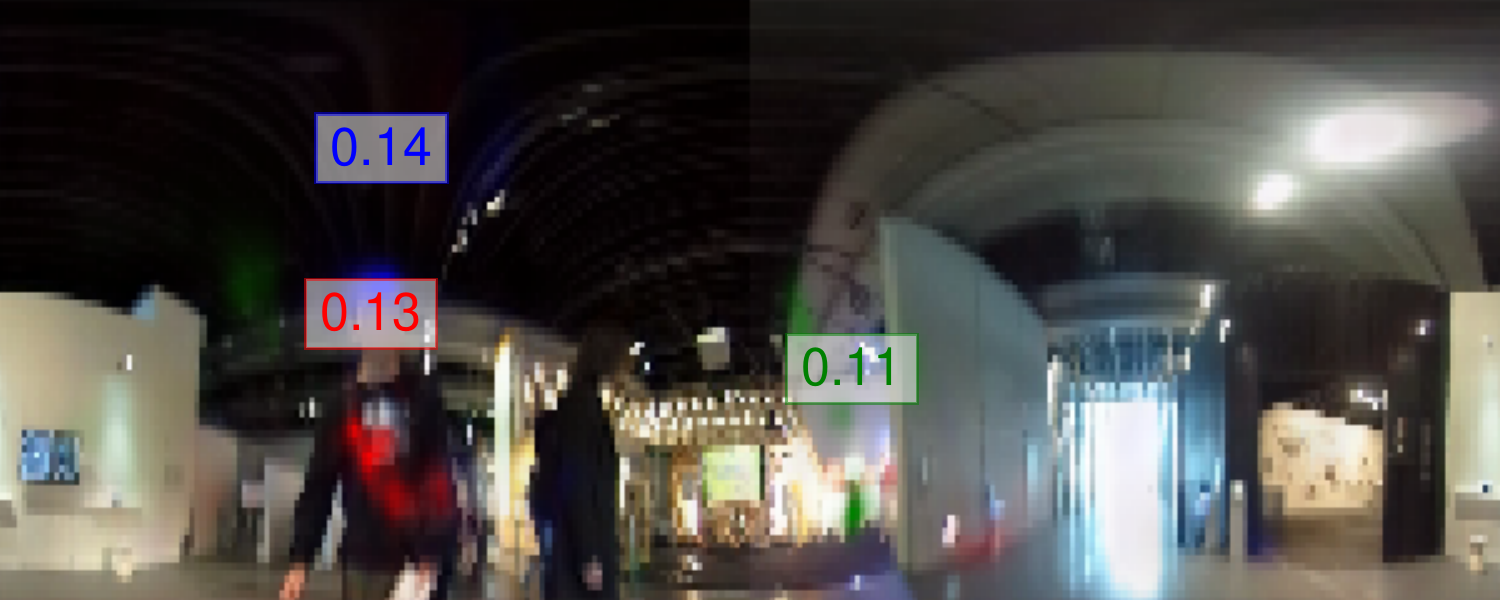} \hfill
    \includegraphics[width=0.325\hsize]{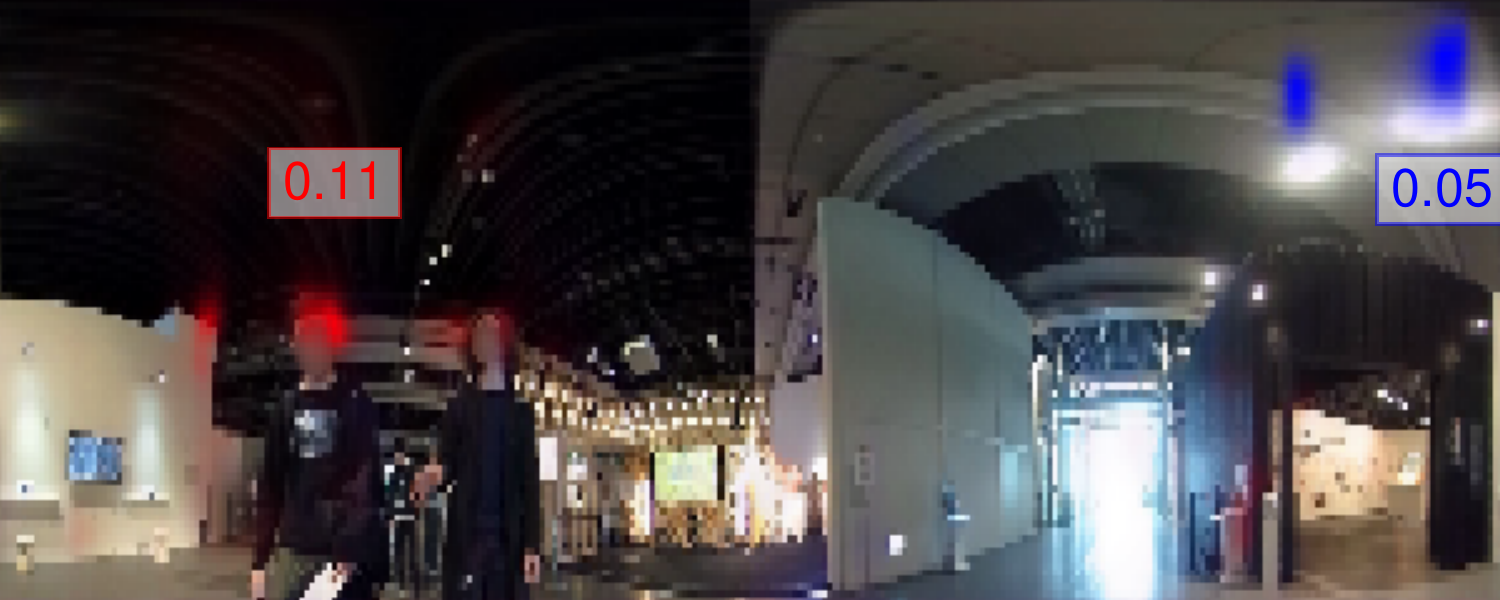} \\
    \vspace{-1.5mm}
    \caption{
        Visualization of localization results for recordings in Miraikan.
        Each row displays excerpts from a single video stream at $2$-second intervals.
        Faces of visitors are blurred for their privacy.}
    \label{fig:miraikan-reult1}
  \vspace{-4mm}
\end{figure*}

This experiment was conducted as a special event in the National Museum of Emerging Science and Innovation (Miraikan)\footnote{https://www.miraikan.jst.go.jp/en/} in Japan.
Peacock \cite{sasaki2017longterm,kanezaki2018goselo} has demonstrated its autonomous navigation on the third floor of the museum (Fig.~\ref{fig:miraikan}), where more than a thousand people visit every day.
We put a $16$-ch microphone array and $360^\circ$ camera on the top of the robot (Fig.~\ref{fig:peacock}-(b)) and recorded audio-visual data of the visitors and exhibits around the robot.
To encourage various reactions of the visitors, we demonstrated SSL by putting an LED tape that indicates DoAs estimated by MUSIC while the robot moved around the floor.
The audio signals were recorded at $16$-kHz and $24$-bit sampling with an A/D converter called RASP-ZX (Systems In Frontier Corp.), and the videos were recorded at $15$ fps and a resolution of $720 \times 1280$ with a $360^\circ$ camera called RICOH THETA S.
The recording was conducted six days in the period from December $2019$ to February $2020$, and the robot has demonstrated seven hours for each day.

Our AV-SSL system was trained from the first four days of the recorded data.
The recordings were split into 1.0-second clips, and we used the top 5000 clips having large sound levels as training data because most of the clips had low sound level.
The hyperparameters of our method were changed from those in Sec.~\ref{sec:evaluation} for adapting the real-world environment.
The hyperparameters $K$, $\alpha_0$, and $\sigma_0$  were set to $8$, $1.0$, and $1.0$, respectively.
The $\epsilon$ was set to 0.1 because the sounds were not stable.

\subsection{Experimental Results}
Fig.~\ref{fig:miraikan-reult1} shows inference examples for the recordings that were not used in the training.
The examples on the top row show that multiple visitors are localized in individual source classes.
These results demonstrate that the proposed method was successfully trained to localize source candidates in a real environment.
In the all examples on the middle row, the right regions were localized.
There was an exhibit producing large mechanical sounds in this region.
Since we took a self-supervised approach, our method was successfully trained to localize such a non-obvious sound source without any supervision.
As shown in the bottom row, our method did not always succeed in localization.
Although the two visitors were successfully localized in the left example of this row, the right visitor was not localized in the other examples, and the left visitor was localized as two source candidates in the middle example.
This problem would be solved by utilizing the continuity of source movements in a video sequence.
Please see the video in the supplementary material that includes not only observed images but also corresponding audio recordings.
\section{Conclusion}

In this paper, we presented a self-supervised training of audio and visual DNNs for AV-SSL using $360^\circ$ images and multichannel audio signals.
Our method trains both DNNs to maximize the ELBO of the probabilistic spatial audio model, which does not require labeled images or source direction as supervision.
Our simulation results confirmed that the trained DNN detected each person separately and determined whether or not each person speaks.
We also demonstrated the applicability of the proposed method by using data recorded in a science museum.
These results indicate the effectiveness of using multichannel audio signals for audio-visual self-supervised learning.
Our future work includes utilizing a video stream to track moving sound source objects.

{\small
\noindent {\bf Acknowledgements}~
The authors would like to thank Mr. Yusuke Date and Dr. Yu Hoshina for their support in the experiment in Miraikan.
This study was partially supported by JSPS KAKENHI No. 18H06490 for funding.}

\bibliographystyle{IEEEbib}

\end{document}